\documentclass[aps,prx,superscriptaddress,reprint]{revtex4-2}
\usepackage{graphicx}
\usepackage{amsmath,amssymb,amsfonts}
\usepackage{caption}
\usepackage{hyperref}
\usepackage{ragged2e}
\usepackage{upgreek}
\usepackage[T1]{fontenc}

\newcommand{\ket}[1]{\ensuremath{\left|#1\right\rangle}}

\newcommand{\tr}[0]{\text{tr}}

\begin{document}

\title[]{Optomechanical quantum bus for donor spins in silicon}

\author{Henri Lyyra}
\email{hesaly@utu.fi}
\affiliation{Department of Physics and Nanoscience Center, P.O. Box 35, University of Jyväskylä, FI-40014 University of Jyväskylä, Finland}
\affiliation{Department of Mechanical and Materials Engineering, University of Turku, Turun yliopisto FI-20014, Finland}

\author{Cliona Shakespeare}
\affiliation{Department of Physics and Nanoscience Center, P.O. Box 35, University of Jyväskylä, FI-40014 University of Jyväskylä, Finland}

\author{Simeoni Ahopelto}
\affiliation{Department of Physics and Nanoscience Center, P.O. Box 35, University of Jyväskylä, FI-40014 University of Jyväskylä, Finland}
\author{Teemu Loippo}
\affiliation{Department of Physics and Nanoscience Center, P.O. Box 35, University of Jyväskylä, FI-40014 University of Jyväskylä, Finland}

\author{Alberto Hijano}
\affiliation{Department of Physics and Nanoscience Center, P.O. Box 35, University of Jyväskylä, FI-40014 University of Jyväskylä, Finland}

\author{Reetu Inkilä}
\affiliation{Department of Physics and Nanoscience Center, P.O. Box 35, University of Jyväskylä, FI-40014 University of Jyväskylä, Finland}
\author{Pyry Runko}
\affiliation{Department of Physics and Nanoscience Center, P.O. Box 35, University of Jyväskylä, FI-40014 University of Jyväskylä, Finland}

\author{Tero T. Heikkil\"a}
\affiliation{Department of Physics and Nanoscience Center, P.O. Box 35, University of Jyväskylä, FI-40014 University of Jyväskylä, Finland}

\author{Juha T. Muhonen}
\email{juha.t.muhonen@jyu.fi}
\affiliation{Department of Physics and Nanoscience Center, P.O. Box 35, University of Jyväskylä, FI-40014 University of Jyväskylä, Finland}

\begin{abstract}
Silicon is the foundation of current information technology, and a promising platform for future quantum information technology as silicon-based qubits exhibit some of the longest coherence times in solid-state. At the same time, silicon is the underlying material for advanced photonics activity, and photonics structures in silicon can be used to define optomechanical cavities where the vibrations of nanoscale mechanical resonators can be probed down to the quantum level with laser light. Here, we propose to bring all these developments together by coupling silicon donor spins into optomechanical structures. We show theoretically and numerically that this allows telecom wavelength optical readout of the spin-qubits and implementing high-fidelity entangling two-qubit gates between donor spins that are spatially separated by tens of micrometers. We present an optimized geometry of the proposed device and discuss with the help of numerical simulations the predicted performance of the proposed quantum bus. We analyze the optomechanical spin readout fidelity and find the optimal donor species for different coupling mechanisms.
\end{abstract}
\
\maketitle

\section{Introduction}

Conventional semiconductor electronics is built on tailoring silicon’s electronic properties by introducing group-V donor atoms. The extensive industrial expertise in this field, coupled with the atomic-scale nature of donor spin qubits, positions them as a natural candidate for the quantum successor to classical bits. Indeed, one of the first scalable quantum computer proposals — the famous Kane quantum computer \cite{kane1998}  — relied on the nuclear spin qubits of individual phosphorus atoms. Ever since Kane's seminal paper, the progress towards a donor spin-based silicon quantum computer has been promising \cite{morello_single-shot_2010,pla_single-atom_2012,pla_high-fidelity_2013}. In isotopically purified silicon, coherence times of 1.1~ms have been demonstrated with the donor electron spin qubits \cite{muhonen2014}, and these can be even further prolonged by dressing the qubits with a microwave drive \cite{laucht2017}. Furthermore, for phosphorus nuclear spin, the coherence time can be prolonged with dynamical decoupling to 30 seconds, a promising demonstration of a quantum memory \cite{muhonen2014}. Single-qubit logical operations with gate fidelities above 99.9~\% have also been demonstrated with the electron spin qubits. \cite{muhonen_quantifying_2015}. In addition to these ion implantation based demonstrations, there has been significant progress with STM lithography approaches \cite{reiner_high-fidelity_2024,edlbauer_11-qubit_2025} and other defects in silicon \cite{higginbottom_optical_2022,higginbottom_memory_2023}.

In Kane's original idea, two nuclear spin qubits would interact using the exchange interaction of two electron spins as a mediator, allowing performing 2-qubit quantum gates with the neighboring qubits. However, in order to get strong enough coupling, the donors have to be implanted with precision of 20 - 30 nm in the original proposal, posing already severe challenges for the integration of the control and measurement electronics for multi-qubit quantum processors. Later, it was realized that the multi-valley nature of silicon gives rise to atomic-scale oscillations in the exchange energy \cite{koiller_exchange_2001}, making the multi-qubit implementation even more challenging.

Despite a recent promising three-qubit demonstration with donor spins \cite{madzik_precision_2022} and entangling two donor spin-qubits \cite{stemp_tomography_2024, stemp_scalable_2025}, a fully scalable interaction between donor spins is yet to be demonstrated, although several theoretical suggestions have been made \cite{tosi_silicon_2017,morse_photonic_2017}. Another challenge currently limiting the scalability of donor spin-based silicon quantum computers is the readout of the qubit states. Although the information could be processed even at 4 K, the reservoir tunneling readout currently in use forces operating at millikelvin temperatures and high magnetic fields \cite{morello_single-shot_2010}.

Here we propose to solve these issues with a new quantum hybrid system coupling the donor spins to an optomechanical platform in order to enable optical readout of the spins as well as a phononic pathway to couple the spins to each other. The platform is inherently scalable as it leverages the existing framework of silicon photonics by transducing the spin information to telecom range photons. In addition, the architecture allows for local coherent spin-spin couplings via the mechanical elements. 
We describe a detailed proposal for silicon based quantum computing utilizing existing optomechanical devices and discuss in detail the usage of the mechanical mode for two-qubit gates. In addition a crucial ingredient for our proposal is the use of a microwave dressed spin state that offers multiple advantages \cite{Hijano2024}.

Optomechanical systems have shown their promise as quantum transducers in multiple different settings. Due to the ability to couple mechanical motion to many different degrees of freedom, there are active projects developing optomechanical platforms for quantum transduction between different quantum systems \cite{laucht_roadmap_2021,barzanjeh_optomechanics_2022}, such as between microwave and optical photons \cite{lambert_coherent_2020} or between defect qubits and optical photons \cite{stannigel_optomechanical_2011}. Information can also be stored in mechanical resonators: There are demonstrations of acoustic systems with quality factors above $10^{10}$ \cite{beccari_strained_2022} and using them as quantum memories for superconducting qubits \cite{wollack2022,lupke2022}. Besides quantum technologies, high fidelity state preparation and readout of a semi-macroscopic mechanical resonator could help answer some very foundational questions about quantum physics, for example testing the continuous spontaneous collapse models \cite{bassi_models_2013,carlesso_present_2022} and gravitational decoherence \cite{bassi_gravitational_2017}. 

In this paper, we propose and analyze the performance of an optomechanical quantum bus that exploits mechanical resonators for optical spin-qubit readout and two-qubit gates. Our proposed platform combines the recent advancements of high-precision single-ion implantation \cite{koenraad_single_2011,donkelaar_single_2015,pacheco_ion_2017,jamieson_deterministic_2017}, high-fidelity control of donor spin-qubits \cite{laucht2017}, and nanofabrication of optomechanical resonators \cite{leijssen_nonlinear_2017}. The donors implanted in isotopically purified silicon couple to an optomechanical sliced photonic crystal nanobeam resonator. The fully silicon on-chip design simplifies the fabrication and scalability compared to architectures that depend on several electrical lines to each qubit. Furthermore, the telecom wavelength optical readout enables quantum information processing at higher temperatures and opens the possibility for connecting the donor spins with photonic communication networks and distributed quantum computing. If the mechanical resonator is cooled down to its ground state, the proposed quantum bus can be also used as a testbed to push the boundaries between the quantum and classical worlds.

The paper is organized as follows: section II gives an overview of the proposed platform, section III discusses spin readout via changes in the mechanical resonance frequency, section IV discusses the spin-spin coupling via the phonon system and section V gives a concrete proposal on the implementation and achievable spin-phonon coupling strengths. Especially we focus on the coherence and fidelity properties as well as the requirements for donor placing which we show are within reach of current implantation technology.

\section{Overview of the hybrid system}

\begin{figure}[t]
\centering
\includegraphics[width=0.99\linewidth]{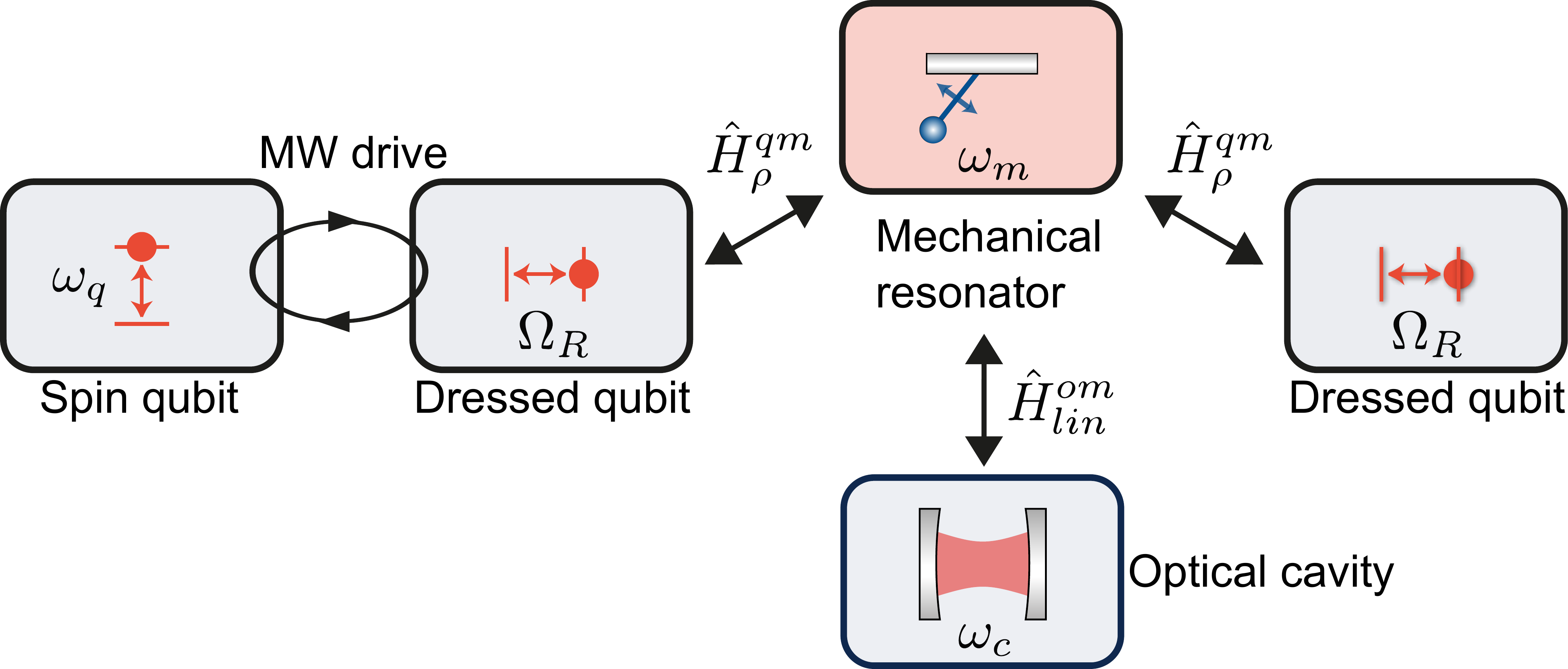}
\captionsetup{justification=Justified}
\caption{Schematic of the optomechanical quantum bus. The dressed qubit's state affects the resonance frequency of the mechanical resonator. The mechanical displacement modulates the resonance frequency of the optical cavity. The spin-dependent mechanical frequency shifts can then be detected by measuring the noise imprinted on the phase quadrature of the light escaping from the cavity. Bringing two spin-qubits near resonance with the resonator allows logical two-qubit operations for spatially separated donor spins. The parameters are discussed in the main text.}
\label{setup}
\end{figure}

Our proposed platform for the optomechanical quantum bus consists of multiple interacting quantum systems, illustrated in Fig.~\ref{setup}. The coupling between the spin and the mechanical resonator can be realized using either a magnetic field gradient or the innate strain coupling. We model the details of the achievable coupling in Sec.~\ref{Sec_couplMecs}. Both mechanisms lead to a mechanical displacement-dependent energy splitting of the qubit levels, i.e., to an interaction Hamiltonian proportional to $\sigma_z$, where $\sigma$ is the Pauli spin matrix (taking $z$ as the quantization axis).  However, to map the qubit spin to the resonator the interaction should involve $\sigma_x$ instead. Moreover, the aim is to achieve active control over the spin-mechanical interaction, such that it can be turned on and off at will. Both requirements can be addressed by applying a microwave drive resonant with the qubit transition frequency \cite{Hijano2024}. This drive ‘dresses’ the qubit \cite{laucht2017}, transforming its eigenstates. In the new dressed basis, the qubit energy splitting is proportional to the microwave drive strength, and hence can be brought in resonance with the mechanical resonance frequencies of typical flexural modes of a micron-scale silicon beams. In this dressed frame the interaction between the mechanical resonator and the spin is proportional to $\sigma_x$.

For the optical side, the coupling is realized through a cavity-optomechanical system: the mechanical movement changes the resonance frequency of an optical cavity. Here, we specifically model the case where we work with an optomechanical system in the non-resolved sideband limit, where the phase of the light escaping the cavity contains information on the instantaneous position of the resonator. We show specific geometries in Sec. \ref{sec_siliconNanoBeams}. The modelled optical wavelength is the optical telecommunication band at 1550 nm.

This system can then be utilized in two different regimes. In the dispersive regime (i.e., when the dressed state is detuned from the mechanical resonator) the spin state can be read out optically by observing changes in the mechanical resonance frequency, analogously on how superconducting qubits are read out via changes in the resonance frequency of a microwave resonator \cite{blais2004}. Notably, this gives a quantum non-demolition readout mechanism. This is in contrast to direct optical readout as used with e.g. NV-centers in diamond. As we can work deep in the dispersive regime, this readout can be accomplished even if the mechanical resonator is in a thermal state.

On the other hand, tuning the spins in resonance with the mechanical resonator would then allow coherent state transfer between the mechanical resonator and the spin. This regime could then be used for both coherent coupling between separate spins via phonons, and to possibly coherently couple the spin state to the light field via the optomechanical system. The effective spin-spin interaction can however also be realized in the dispersive regime (where qubits are resonant but the mechanical resonator is detuned), which can be advantageous if there is a small but finite thermal occupancy of the mechanical mode. This is discussed in detail in Sec.~\ref{sec_spinspin}.

\subsection{Dressed frame coupling}
\label{sec:dressedHam}

Here we present the Hamiltonian formulation of coupling the spin qubit and the mechanical resonator. For clarity, we neglect the hyperfine interaction between the donor bound electron spin and the donor nuclear spin. The qubit system here is the donor bound electron spin.

We start with the canonical spin qubit Hamiltonian and assume a static magnetic field $B_0$ is applied. The energy levels of the electron spin states are separated by the Zeeman splitting $\gamma_e B_0$, where $\gamma_e := \frac{g_e \mu_B}{\hbar}$ is the electron gyromagnetic ratio, $g_e$ is the electron $g$-factor, $\mu_B$ is the Bohr magneton. If we then add a mechanical resonator whose displacement $\hat{a}^\dagger + \hat{a}$ modulates the Zeeman splitting, the Hamiltonian of the total qubit-resonator system reads
\begin{align}\label{qubitOsc}
\frac{ \hat{H}^{qm} }{\hbar} = \frac{1}{2}\left[ \gamma_e B_0 \hat{\sigma}_z + 
2 \omega_m \hat{a}^\dagger \hat{a} + \lambda \hat{\sigma}_z\left( \hat{a}^\dagger + \hat{a} \right) \right]\,,
\end{align}
where we have shifted the mechanical resonator's ground state energy to zero by omitting the zero-point fluctuation term, $\omega_m$ is the mechanical resonance frequency, and $\hat{a}^\dagger$ and $\hat{a}$ are the creation and annihilation operators for the mechanical resonator mode phonons, respectively.  
The coupling strength $\lambda$ between the resonator and the spin qubit depends on the coupling mechanism, donor species, and device design, see Sec.~\ref{Sec_couplMecs} for details.

The spin qubit states can be controlled by applying an AC electromagnetic field (drive in the following) with a magnetic component orthogonal to the applied DC field ($\Vec{B}_1\perp\vec{B}_0$) oscillating at the MW frequency $\nu_{MW}$. The Hamiltonian with the oscillating magnetic field (which we model as a classical drive) can be written as 
\begin{align}\label{dressedQubitH0}
\begin{aligned}
    \frac{\hat{H}^{qm}}{\hbar} = \frac{1}{2} 
    & \left[ \gamma_e B_0 \hat{\sigma}_z + \gamma_e B_1 \cos\left(\nu_{MW}t \right)\hat{\sigma}_x \right. \\
    & \left. + 2 \omega_m \hat{a}^\dagger \hat{a} + \lambda \hat{\sigma}_z\left( \hat{a}^\dagger + \hat{a} \right) \right]\,.
\end{aligned}
\end{align}
When $\nu_{MW}$ is in resonace with the Zeeman splitting, the drive starts driving the spin transition, causing coherent oscillations between the states $\ket{\downarrow}$ and $\ket{\uparrow}$ at the Rabi frequency $\Omega_R := \frac{1}{2}\gamma_e B_1$. Here, $\ket{\downarrow}$ and $\ket{\uparrow}$ are the eigenstates of $\hat{\sigma}_z$.

By moving into the coordinates rotating with the MW drive and applying a change of basis, 
%$\ket{\downarrow} \rightarrow \ket{+} := \frac{1}{\sqrt{2}}( \ket{\downarrow} + \ket{\uparrow} ),\,\ket{\uparrow} \rightarrow \ket{-} := \frac{1}{\sqrt{2}}( \ket{\downarrow} - \ket{\uparrow} )$,
Eq.~\eqref{dressedQubitH0} simplifies into
\begin{align}\label{dressedQubitH}
\frac{ \hat{H}^{qm}_{\rho} }{\hbar} = \frac{1}{2}\left[ 
    \Omega_R \hat{\sigma}_z + \Delta \nu \hat{\sigma}_x  
    + 2 \omega_m \hat{a}^\dagger \hat{a} - \lambda \hat{\sigma}_x\left( \hat{a}^\dagger + \hat{a} \right) \right]\,,
\end{align}
where the original eigenstates of the bare qubit $\ket{\downarrow}$ and $\ket{\uparrow}$ are now the eigenstates of $\hat{\sigma}_x$ in these coordinates, and we have denoted the MW detuning as $\Delta\nu := \nu_{MW} - \gamma_e B_0$. 

If the qubit is driven resonantly, $\Delta\nu = 0$, we get a new qubit, \textit{the dressed qubit} with $\Omega_R$ as its eigenfrequency. Previous experiments with silicon donor spins have shown that the dressing detaches the qubit from the low frequency noise, and can prolong the coherence times by one order of magnitude giving coherence times fo $T_{2\rho}^{\star} = 2.4$ ms, and $T_{2\rho}^{Hahn} = 9$ ms for a $^{31}$P donor in silicon \cite{laucht2017}. Importantly, since the eigenfrequency $\Omega_R$ is tuned by the intensity of the MW drive, the qubit can be brought into and out of resonance with a mechanical resonator by tuning the microwave drive, hence giving precise control of the interaction time between the qubit and the mechanical resonator. Also, the Bloch vectors corresponding to the eigenstates of the dressed qubit are perpendicular to those of the bare spin-qubit, which allows for the transverse coupling with the mechanical resonator. 

Finally, if $\Delta\nu = 0$ and we apply the rotating wave approximation (RWA), we have the standard Jaynes-Cummings Hamiltonian between the dressed qubit and the mechanical resonator
\begin{align}\label{dressedQubitH1RWA}
\frac{ \hat{H}^{qm}_{\rho ,RWA} }{\hbar} = \frac{1}{2}\left[ 
    \Omega_R \hat{\sigma}_z 
    + 2 \omega_m \hat{a}^\dagger \hat{a} - \lambda \left( \hat{\sigma}_-\hat{a}^\dagger + \hat{\sigma}_+\hat{a} \right) \right]\,.
\end{align}

In the numerical simulations, we omit the RWA and use the Hamiltonian of Eq.~\eqref{dressedQubitH} to avoid the influence of any unnecessary assumptions in our analysis. However, the RWA Hamiltonian in Eq.~\eqref{dressedQubitH1RWA} combined with other assumptions is a useful tool to provide intuition on why and how the quantum bus implements qubit readout and two-qubit gates, as we discuss in Sec.~\ref{sec_readout} and Sec.~\ref{sec_spinspin}, respectively.

\subsection{The optomechanical resonator}

For completeness, we describe here shortly the optomechanical interaction. Although it is not the main component in the modeling we present, it is crucial to understand what is the measured quantity in our readout. 

The key ingredient of our proposed quantum bus is the ability of mechanical resonators to interact simultaneously with a plethora of different physical systems via changes of displacement, strain, or changes in the device geometry. Especially the field of optomechanics relies on  the mechanical motion of a resonator modifying the resonance frequency $\omega_c$ of an optical cavity. This optomechanical interaction can among other things be exploited to monitor and control the mechanical oscillator, which is an integral part of the proposed optomechanical quantum bus. 
In coordinates rotating at the laser frequency $\omega_L$ and linearizing the optomechanical interaction term, the optomechanical Hamiltonian can be written in the form
\begin{align}\label{omLinH}
\begin{aligned}
\frac{ \hat{H}^{om}_{lin} }{\hbar} =
& \omega_m \hat{a}^\dagger \hat{a} - \Delta_{CL} \delta \hat{b}^\dagger \delta\hat{b} \\
&- g_0 \sqrt{ \langle n_{cav} \rangle } \left( \hat{a}^\dagger + \hat{a} \right)  \left( \delta \hat{b}^\dagger + \delta  \hat{b} \right) 
\end{aligned}
\end{align}
where $\Delta_{CL} = \omega_c - \omega_L$ is the detuning between the optical cavity and the laser drive, $\delta\hat{b}^\dagger$ and $\delta\hat{b}$ are the photon number fluctuation creation and annihilation operators, $g_0$ is the single photon optomechanical coupling strength, and $\langle n_{cav} \rangle$ is the average photon number inside the cavity \cite{aspelmeyer_cavity_2014}. The interaction term in Eq.~\eqref{omLinH} couples the displacement operator of the mechanical resonator to the phase quadrature of the photon field. This imprints the mechanical motion to the phase noise of the light leaking out of the optical cavity. Specifically our observable is the symmetrized power spectral density of the mechanical motion \cite{bowen_quantum_nodate}, as we detail below.

In what follows, we mostly neglect the optomechanical interaction, and  concentrate only on the spin-mechanics part of the optomechanical quantum bus, described by the Hamiltonian in Eq.~\eqref{dressedQubitH}. In the next two sections, we discuss how the described quantum bus can be used for optomechanical readout and controlled two-qubit gates for donor spin qubits in silicon.
\section{Spin readout via changes in the mechanical resonance frequency }\label{sec_readout}

\subsection{Dispersive limit}

To get intuition of how and why the optomechanical spin-qubit readout works, let us first consider the qubit-mechanics Hamiltonian under the RWA in Eq.~\eqref{dressedQubitH1RWA}. Now, we consider the \textit{dispersive regime} where the qubit and the mechanical resonator do not exchange energy. It takes place when $\lambda \ll 2 |\Omega_R - \omega_m| =: 2 | \Delta_R |$ and $n_{\rm th} := \langle \hat{a}^\dagger \hat{a} \rangle \ll \frac{ \Delta_R^2 }{ \lambda^2 }$ \cite{boissonneault_dispersive_2009}. In this limit, the Hamiltonian can be approximated in rotating coordinates as 
\begin{align}\label{Hdisp}
    \frac{H_{disp}^{qm}}{\hbar} = \left( \omega_m + \frac{\lambda^2}{4 \Delta_R}\hat{\sigma}_z \right) \hat{a}^\dagger \hat{a} + \frac{1}{2} \Omega_R \hat{\sigma}_z + \frac{\lambda^2}{8 \Delta_R} \sigma_z\,.
\end{align}

If the bare qubit was initialized to the state $\ket{\uparrow}$ ($\ket{\downarrow}$), then after adiabatic dressing the qubit will be in the eigenstate $\ket{+}$ ($\ket{-}$) of the dressed qubit energy operator $\hat{\sigma}_z$ in Eq.~\eqref{Hdisp}. The $\hat{a}^\dagger\hat{a}$ term in Eq.~\eqref{Hdisp} illustrates that in the dispersive regime, the mechanical resonance frequency gets shifted by $\pm\frac{\lambda^2}{4\Delta_R}$ where "+" ("-") corresponds to the spin-qubit's initial states $\ket{\uparrow}$ ($\ket{\downarrow}$) before dressing it. Optically measuring the shift of the mechanical frequency allows readout for the spin state via the mechanical resonator similarly to how monitoring the resonance frequency of a waveguide resonator is used for reading out the coupled superconducting qubit \cite{blais2004,wallraff2004}. Furthermore, the interaction term $\frac{\lambda^2}{4 \Delta_R}\hat{\sigma}_z \hat{a}^\dagger \hat{a}$ in Eq.~\eqref{Hdisp} commutes with the local Hamiltonians of the qubit and the mechanical resonator, and thus the readout is a non-demolition measurement.

Note also that the criterion for the dispersive regime $\langle \hat{a}^\dagger \hat{a} \rangle \ll \frac{ \Delta_R^2 }{ \lambda^2 }$ effectively sets the temperature range of the readout: Higher expectation value for the phonon number means that larger detuning must be used to satisfy the condition. Increasing the detuning, in turn, diminishes the expected frequency shift. Otherwise the thermal state of the resonator here does not necessarily matter as in the dispersive regime it does not lead to the thermalization of the qubit state.

\begin{figure}
\centering
\includegraphics[width=0.99\linewidth]{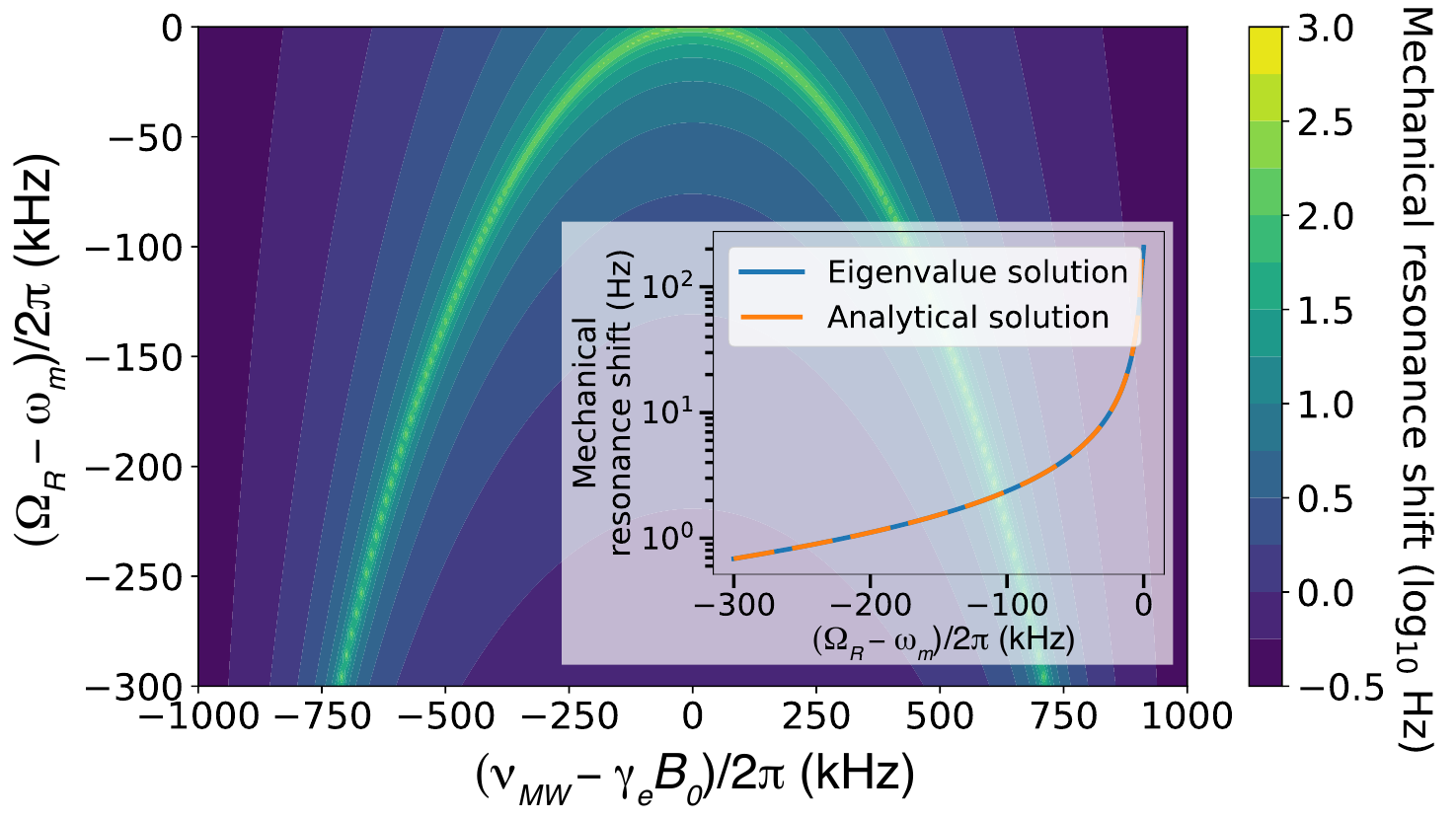}
\captionsetup{justification=Justified}
\caption{The expected shift of the mechanical resonance frequency as a function of the detunings. These shifts are calculated from the energy levels of the qubit-oscillator Hamiltonian, and thus do not consider any finite linewidths of the qubit or the resonator. The results show an interplay between the detunings: the maximal shift follows well the qubit-mechanics resonance condition $\omega_m = \sqrt{\Omega_R^2 + \Delta \nu^2}$.
The inset shows a linecut at $\nu_{MW}=\gamma_eB_0$, together with an analytical calculation for the expected shift $\Delta \omega_m = \lambda^2 \Omega_R / [ 2(\omega_m^2 - \Omega_R^2) ]$ \cite{noRWAanalyticalShift}. For the numerical values here, we have fixed the coupling strength as $\lambda/(2\pi) = 1$ kHz, and $\omega_m/(2\pi) = 1$ MHz.}
\label{eigenvalShifts}
\end{figure}

\subsection{Numerical simulation}

The approximated Hamiltonian in Eq.~\eqref{Hdisp} assumes both the RWA and dispersive limit and does not include any effects of finite linewidths. Below we numerically benchmark the system in a more realistic noisy environment and test how much compromising the RWA and dispersive assumptions affect the readout. 
We relax any assumptions concerning RWA and the dispersive limit in the model and apply the general qubit-mechanics Hamiltonian in Eq.~\eqref{dressedQubitH}. We restrict here to the regime where $\Omega_R \leq \omega_m$.

Figure \ref{eigenvalShifts} illustrates the shift of the energy level spacing of the qubit-mechanics Hamiltonian in Eq.~\eqref{dressedQubitH} as a function of the Rabi and Zeeman detunings. The calculation shows that when the natural mechanical resonance frequency $\omega_m$ is comparable to the driven qubit's eigenfrequency, or in other words $\omega_m \approx \sqrt{\Omega_R^2 + \Delta \nu^2}$, we expect a shift of the mechanical resonance frequency that depends on the spin state.
The eigenvalue calculation shows also that the shift is maximized in resonant driving, namely when $\Delta \nu = 0$ and $\Omega_R \approx \omega_m$. 

Next, we consider the dynamics of the system including finite linewidths for the qubit and the mechanical resonator and noting that the mechanical resonator might initially be in a thermal state.  
We use the Python Qutip \cite{qutip1,qutip2} package to numerically study the relation between the mechanical frequency shift, the detunings and the coupling strength $\lambda$ using Eq.~\eqref{dressedQubitH} to model the dynamics. 
The aim of the simulations is to study the spin-mechanics coupling strength needed for successful spin readout via the optomechanical quantum bus and to aid in analyzing the performance of different coupling mechanisms and designs. Here we used a 100 Hz shift of the resonance frequency shift as the benchmark value as it is a usual value for the mechanical linewidth in the measured devices. We do note that the 100 Hz shift discussed here is a stringent requirement, as both smaller shifts than the linewidth can be measured (e.g. with phase locked-loops or using optomechanical self-oscillations to narrow the mechanical linewidth \cite{forceSensingSelfOscillations}) and also because the linewidth could be improved using e.g. phononic shielding \cite{chan_laser_2011}.

In the numerical simulations, the linewidths are implemented by introducing decay channels for the qubit and mechanical resonator with the Gorini-Kossakowski-Sudarshan-Lindblad (GKSL) master equation
\begin{align}\label{gksl}
 \frac{d\hat{\rho}}{dt} =& i \left[ \hat{\rho}, \hat{H}^{qm}_{\rho} \right] +\\& \sum_{k = 1}^5 \kappa_k \left[ \hat{A}_k \hat{\rho} \hat{A}^\dagger_k -\frac{1}{2} \left( \hat{A}^\dagger_k \hat{A}_k \hat{\rho} + \hat{\rho} \hat{A}^\dagger_k \hat{A}_k  \right) \right]\nonumber \,,
\end{align}
where $\hat{A}_k$ are the jump operators corresponding to the decay channels and $\kappa_k$ are their decay rates. 
In our case, $\hat{\rho}$ is the density matrix of the total qubit--mechanics system, the jump operators are $\hat{A}_1 = \sqrt{(n_\mathrm{th}+1)} \hat{a}$, $\hat{A}_2 =\sqrt{n_\mathrm{th}}\hat{a}^\dagger$ for the dissipation and heating of the mechanical resonator, $\hat{A}_3 = \hat{\sigma}_-$, $\hat{A}_4 = \hat{\sigma}_+$ for the qubit $T_1$ process, and $\hat{A}_5 = \hat{\sigma}_z$ for the qubit pure dephasing term, and the unitary part is governed by the dressed qubit--phonon Hamiltonian $\hat{H}^{qm}_{\rho}$ in Eq.~\eqref{dressedQubitH}. The decay rates $\kappa_k$ are chosen so that the linewidths of the resulting noise spectral densities of the uncoupled dressed qubit $T_1$ and $T_2$ processes and the mechanical resonator correspond to their measured values in the practical implementation we describe in Sec.~\ref{sec_siliconNanoBeams}; $1/T_1 = 10$ Hz \cite{laucht2017}, $1/T_2 = 100$ Hz \cite{laucht2017}, and mechanical linewidth of $100$ Hz. We have used equal relaxation and excitation rates $(\hat{\sigma}_-$, $\hat{\sigma}_+$) as the origin for these processes for the dressed qubit is the electromagnetic environment, which acts essentially like an infinite temperature bath for these frequencies \cite{muhonen2014,laucht2017}.

From the solution of the master equation Eq.~\eqref{gksl}, we calculate the symmetrized spectral density $\bar{S}_{\hat{X}\hat{X}}(\omega) := [S_{\hat{X}\hat{X}}(\omega) + S_{\hat{X}\hat{X}}(-\omega)]/2$. Here 
\begin{align}
S_{\hat{X}\hat{X}}(\omega) = \int_{-\infty}^{+\infty} \langle \hat{X}(t)\hat{X}(0) \rangle e^{i\omega t} dt\,,
\end{align}
is the power spectral density of the mechanical resonator, $\hat{X} = \left( \hat{a}^\dagger + \hat{a} \right)$ is the displacement operator of the mechanical resonator, and $\langle \rangle$ corresponds to the expectation value. We determine the shifted mechanical frequency from $\bar{S}_{\hat{X}\hat{X}}(\omega)$ for different coupling strengths $\lambda$ and detunings by fitting a Lorentzian lineshape to the calculated spectra. 

\begin{figure}
\captionsetup{justification=Justified}
\includegraphics[width=0.98\linewidth]{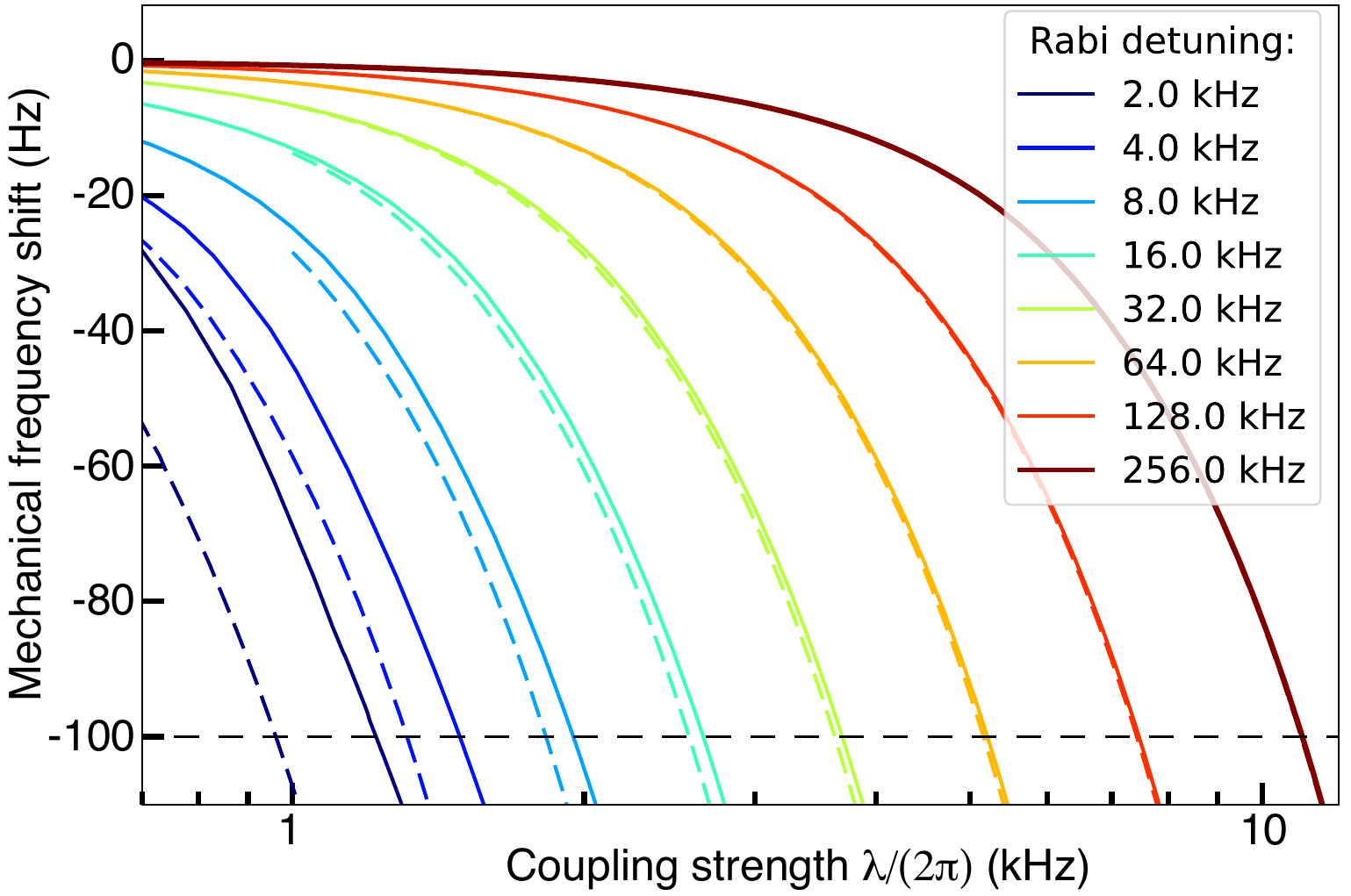}
\caption{Shift of the mechanical resonance frequency as a function of the coupling strength for different detunings between the mechanical resonance and the qubit's Rabi frequency. This detuning value effectively defines the possible operation temperature. 
Here, $\omega_m/(2\pi) = 1$~MHz and $\Delta \nu = 0$. The horizontal black line highlights the 100 Hz shift. 
The solid and dashed lines correspond to the mechanical resonator initialized in the thermal state ($n_{\rm th} = 3)$ and ground state, respectively.}
\label{exampleShiftsQiskit}
\end{figure}

An example plot of the calculated mechanical frequency shifts is presented in Fig.~\ref{exampleShiftsQiskit}. 
We have done the simulation with the mechanical resonator ($\omega_m/(2\pi) = 1$~MHz) both in the ground state and with a small thermal population $n_{\rm th} = 3$. We see that the results for these two starting states differ for small Rabi detunings, highlighting the fact that a thermal state does not prevent the spin readout protocol, but it increases the required Rabi detuning to satisfy the dispersive regime condition. We note that it is obviously very hard if not impossible to directly cool a 1 MHz resonator to these low thermal populations, but very low phonon populations can be achieved with feedback cooling methods \cite{guo_feedback_2019,kumar_single-laser_2022}. The numerical values are still mostly chosen for numerical convenience to highlight how the protocol will scale with the number of phonons, specifically showing the need for reaching the dispersive limit that scales with the phonon number.

\begin{figure}
\captionsetup{justification=Justified}
\includegraphics[width=0.98\linewidth]{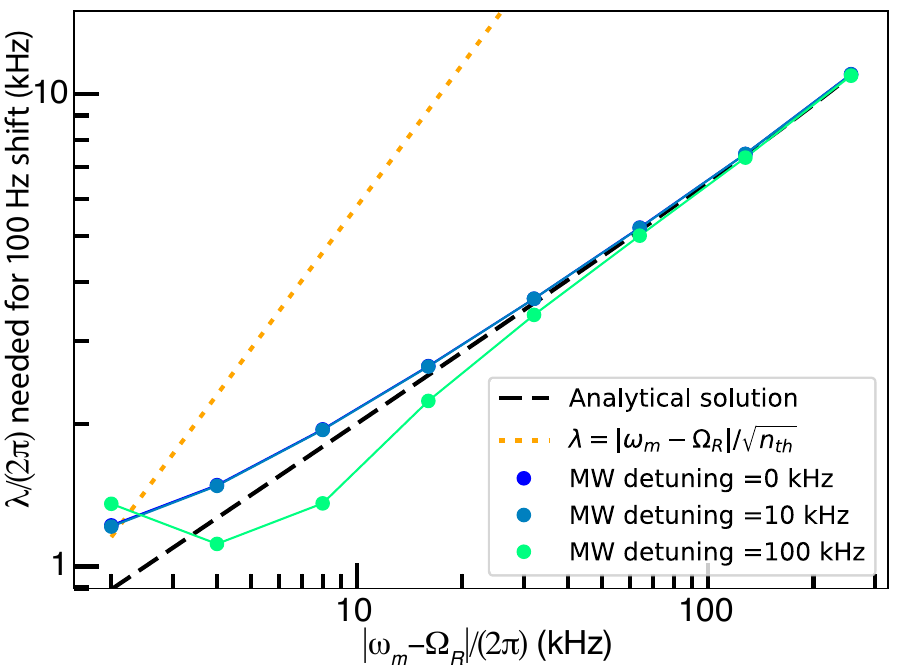}
\caption{Required coupling strength $\lambda$ to cause the 100 Hz shift of the mechanical frequency for different combinations of MW and Rabi detunings. Dashed black line shows the expected Jaynes-Cummings behaviour $\Delta \omega_m = \lambda^2 \Omega_R / [ 2(\omega_m^2 - \Omega_R^2) ]$ \cite{noRWAanalyticalShift}. Dotted yellow line marks roughly the dispersive limit boundary. Here $n_{\rm th} = 3$. The two lowest detuning curves lie on top of each other on this scale.
}
\label{200hzqiskit}
\end{figure}

We repeated the simulation for different values of MW detuning to study its effect on the readout. Figure \ref{200hzqiskit} shows the crossover values leading to the mechanical frequency shift of 100 Hz,  the magnitude of the mechanical linewidth, for each detuning combination for the initial state with $n_{\rm th} = 3$. At high Rabi detunings we see consistency with the simple eigenstate calculation of Fig.~\ref{eigenvalShifts} for both $n_{\rm th} = 0$ and $n_{\rm th} = 3$ and all MW detunings.

There is no notable difference between the resonant MW drive and the 10 kHz MW detuning. As expected, the results deviate notably from the prediction of the dispersive regime for small Rabi detunings in all cases.  For a moderate MW detuning of 100 kHz we see that the results do not match the resonant case. Interestingly, at intermediate Rabi detunings the 100 Hz mechanical frequency shift is now reached at smaller $\lambda$. This is in line with the eigenvalue simulation in Fig.~\ref{eigenvalShifts} which reflects the combined roles of $\Omega_R$ and $\Delta \nu$ in the driven qubit's eigenfrequency and thus its resonance conditions with the mechanical resonator. This can be useful in cases where inducing high Rabi frequency has undesired effects, such as sample heating caused by the high MW power the large $B_1$ requires: detuning the MW drive allows more pronounced mechanical shift at lower $\Omega_R$ without needing a larger $\lambda$ which is harder to tune. We also see a good match between the analytical solution at the dispersive limit and the simulated full dynamics at higher Rabi detunings for both the small and the moderate MW detunings.
\section{Spin-spin coupling via the phonon system}\label{sec_spinspin}

Exploiting an intermediary quantum system has proven useful for implementing 2-qubit gates for other physical implementations of a qubit: Two double quantum dots were coupled via MW line integrated on the chip in \cite{borjans2020}, coupling two superconducting qubits via an MW transmission line was experimentally demonstrated in \cite{sillanpaa2007,majer2007}, and entangling qubits via a mechanical oscillator has been considered theoretically for NV centers in diamond in \cite{{rosenfeld2021}} and similarly for atoms in optical cavities in \cite{sorensen2003}. Next, we will discuss how it applies to our system. 

\subsection{Dispersive limit}

\begin{table}
    %\centering
    \captionsetup{justification=Justified}
    \begin{tabular}{ccl}
      \ket{\downarrow \downarrow }   & $\xrightarrow{\sqrt{iSWAP}}$ & \ket{\downarrow \downarrow } \\
      \ket{\downarrow \uparrow }   & $\xrightarrow{\sqrt{iSWAP}}$ & $\frac{1}{\sqrt{2}}\left( \ket{\downarrow \uparrow } + i \ket{\uparrow \downarrow } \right)$ \\
      \ket{\uparrow \downarrow }   & $\xrightarrow{\sqrt{iSWAP}}$ & $\frac{i}{\sqrt{2}}\left( \ket{\downarrow \uparrow } - i \ket{\uparrow \downarrow } \right)$ \\
      \ket{\uparrow \uparrow }   & $\xrightarrow{\sqrt{iSWAP}}$ & \ket{\uparrow \uparrow } \\
    \end{tabular}
    \caption{The two-qubit gate $\sqrt{iSWAP}$ turns the product states $\ket{\downarrow \uparrow }$ and $\ket{\uparrow \downarrow }$ into maximally entangled states. In combination with single qubit unitary operations, $\sqrt{iSWAP}$ forms the universal gate set.}
    \label{SQiSW_table}
\end{table}

Let us consider what happens when two donor spin qubits are coupled to the same mechanical resonator but do not interact directly with each other. The total qubit-mechanics-qubit Hamiltonian can be written in the dressed frame as
\begin{align}\label{dressedQubitH2}
\frac{ \hat{H}^{qmq}_{\rho} }{\hbar} = \omega_m \hat{a}^\dagger \hat{a} + \frac{1}{2} \sum_{k = 1}^2  \Omega_R^k \hat{\sigma}_z^k + \Delta \nu^k \hat{\sigma}_x^k - \lambda^k \hat{\sigma}_x^k\left( \hat{a}^\dagger + \hat{a} \right) \,,
\end{align}
where the superscript $k \in \{1, 2\}$ labels the qubits 1 and 2.

Now, to provide intuition on how and why the quantum bus can implement entangling 2-qubit gates, we follow the approach presented in \cite{blais2004,blais2007}: By performing the RWA and taking the dispersive limit $\lambda^k \ll 2 |\Omega_R^k - \omega_m|$, we can approximate the combined Hamiltonian as
\begin{align}\label{dressedQubitH2disp}
\begin{aligned}
    \frac{ \hat{H}^{qmq}_{disp} }{\hbar} =& \left[ \omega_m + \sum_{j=1}^2 \frac{ (\lambda^j)^2 }{4 (\Omega_R^j - \omega_m)  } \sigma_z^j \right] \hat{a}^\dagger \hat{a} \\
    &+ \frac{1}{2} \sum_{k = 1}^2  \left[ \left( \Omega_R^k + \frac{(\lambda^k)^2}{4(\Omega_R^k-\omega_m)} \right) \hat{\sigma}_z^k + \Delta \nu^k \hat{\sigma}_x^k \right] \\
    &+ \frac{\lambda^1 \lambda^2 ( \Omega_R^1 - \omega_m + \Omega_R^2 - \omega_m )}{8( \Omega_R^1 - \omega_m) ( \Omega_R^2 - \omega_m )} \left( \hat{\sigma}_+^1 \hat{\sigma}_-^2 + \hat{\sigma}_-^1 \hat{\sigma}_+^2 \right)\,.
    %- \lambda^k \hat{\sigma}_x^k\left( \hat{a}^\dagger + \hat{a} \right) \,,
\end{aligned}
\end{align}

In this noiseless dispersive case, the interaction term of the Hamiltonian Eq.~\eqref{dressedQubitH2} results in the $\sqrt{iSWAP}$ gate between the two spin-qubits when the interaction time is $\frac{ ( \Omega_R^1 - \omega_m )( \Omega_R^2 - \omega_m )}{\lambda_1 \lambda_2 \left( \Omega_R^1 - \omega_m + \Omega_R^2 - \omega_m \right)} $ . Table \ref{SQiSW_table} shows how $\sqrt{iSWAP}$ operates on the two-qubit basis states. The gate can be used to entangle the qubits and forms a universal gate set with unitary one-qubit gates.

\subsection{Numerical simulation}

\begin{figure}[t]
\captionsetup{justification=Justified}
\includegraphics[width=0.90\linewidth]{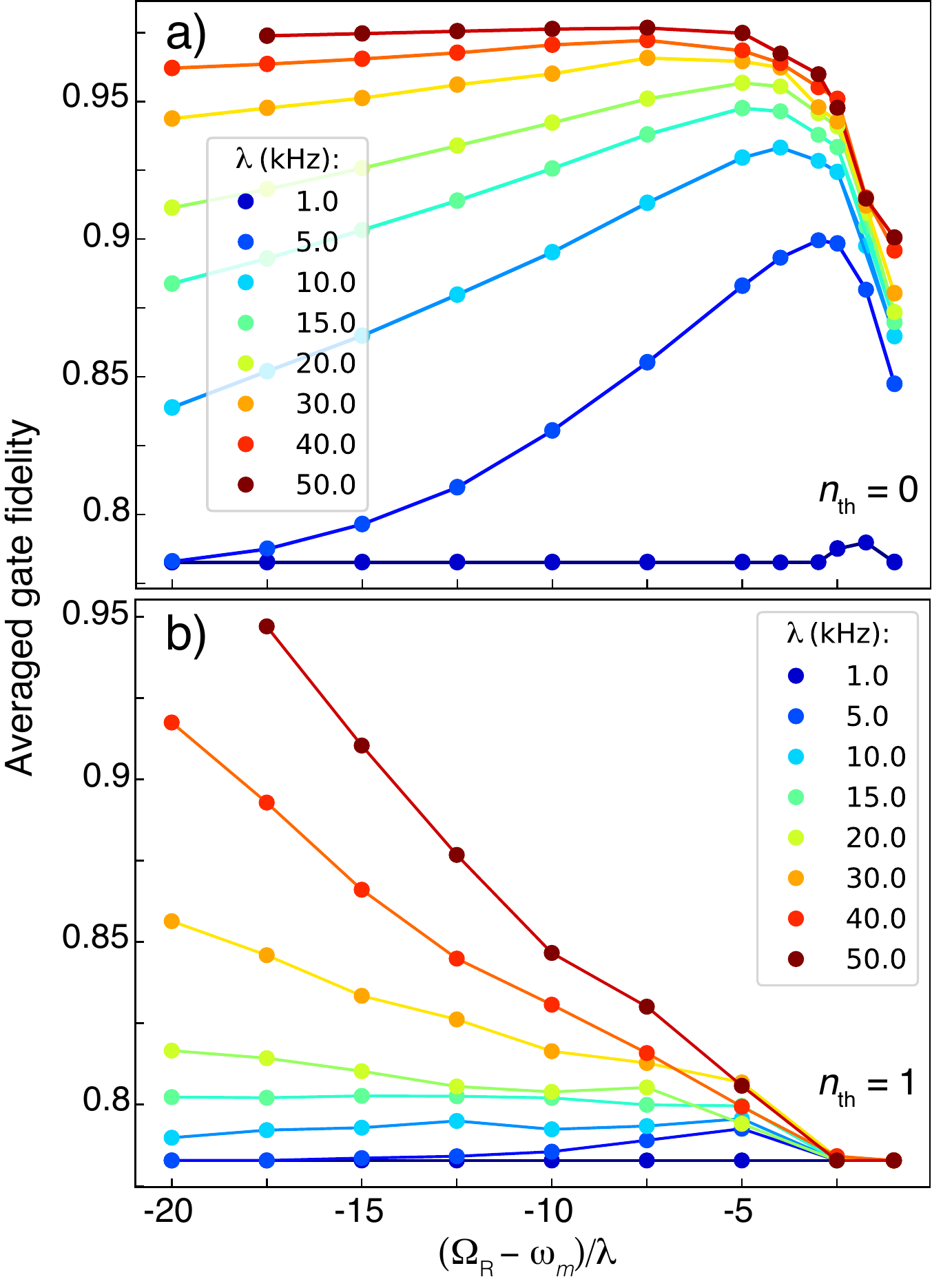}
\caption{The $\sqrt{iSWAP}$-gate fidelity as function of Rabi detuning and coupling strength.}
\label{gateFid_groundState}
\end{figure}

Again, we perform numerical simulations to estimate how our quantum bus performs in a more realistic noisy environment. As in Sec.~\ref{sec_readout}, we relax all the assumptions concerning the dispersive limit and RWA to avoid their unnecessary influence on the results and use the Hamiltonian of Eq.~\eqref{dressedQubitH2}.

We consider the situation where the dressed qubits are in resonance $\Omega_R^1 = \Omega_R^2 = \Omega_R$ and their coupling strengths are equal $\lambda_1 = \lambda_2  = \lambda$. In this case the dispersive regime's ideal interaction time to implement the $\sqrt{iSWAP}$ gate becomes $\frac{1}{2 \lambda}\frac{\Omega_R - \omega_m}{\lambda} $. We see two competing roles for the parameters: the larger the $\lambda$ or the $\frac{\lambda}{\Omega_R - \omega_m} $ is, the faster the gate becomes. Simultaneously, increasing either of these terms pushes us further away from the dispersive regime $\lambda \ll 2 |\Omega_R - \omega_m|$, and the assumptions leading to the approximated Hamiltonian in Eq.~\eqref{dressedQubitH2disp} are not valid anymore. This emphasizes the importance of the Rabi detuning in implementing the 2-qubit gate under realistic noisy conditions: since the Markovian decoherence of the qubits in Eq.~\eqref{gksl} leads to monotonic loss of information, quick gate implementation is desirable. On the other hand, only increasing the coupling strengths violates the underlying assumptions and hinders the gate implementation. Therefore, the Rabi-detuning must be increased to remain in the dispersive regime while increasing $\lambda$ to make the gate fast enough to beat the decoherence.

We use again the GKSL master equation to implement the decoherence processes for the two qubits and the mechanical resonator to simulate the dynamics of the two-qubit system like in Sec.~\ref{sec_readout}. We use two (virtual) ancillary qubits that don't interact with each other or the other systems to construct the Choi matrix of the original 2-qubit system by using the Choi–Jamiołkowski isomorphism. The Choi matrix fully captures the implemented two-qubit gate for any input state and it allows a straightforward way to calculate the averaged gate fidelity between the target gate -- $\sqrt{iSWAP}$ -- and the dynamical map which we get as the numerical solution of the master equation. 

The dynamical map consists of a time-parametrized family of quantum channels $\Phi_t$ for all time points $t$ of the simulation. The averaged gate fidelity is defined as the quantity
\begin{align}
    \mathcal{F}_{\Phi_t,\,\mathcal{U}} ( \phi ) := \left( \tr \left[ \sqrt{ \sqrt{\Phi_t (\phi) } \, \mathcal{U}(\phi) \, \sqrt{\Phi_t (\phi) } } \right] \right)^2
\end{align}
averaged over all initial states $\phi$. In \cite{magesan_gate_2011} it was shown that the averaged gate fidelity can be written as 
\begin{align}\label{chanFidKraus}
    \mathcal{F}_{\Phi,\,\mathcal{U}} = \frac{ d + \sum_n \tr\left[ K_n U^\dagger \right] \tr\left[ K_n^\dagger U \right] }{ d^2 + d }\,.
\end{align}
Here $\mathcal{U}(\phi) = U\phi\, U^\dagger$ is the target unitary gate to be implemented, $d$ is the Hilbert space dimension of the input and output of the implemented channel $\Phi$, and $\{K_n\}_n$ is any set of Kraus operators for $\Phi$ such that $\Phi(\phi) = \sum_n K_n \phi K_n^\dagger$ for any initial state $\phi$. After constructing the Choi matrix, we convert it into a corresponding set of Kraus operators and calculate the channel fidelity as in Eq.~\eqref{chanFidKraus}. The same is repeated for all interaction times and detuning-coupling strength combinations.

In Fig. \ref{gateFid_groundState}(a) we present the simulated gate fidelity with different coupling strengths and Rabi detunings with the resonator initially in the ground state, and in Fig. \ref{gateFid_groundState}(b) with the resonator initialized to one thermal phonon population. The gate fidelity value is picked at the optimal interaction time found separately in the simulation for each detuning-coupling pair. 

As expected, the non-zero qubit and resonator linewidths and the deviation from the dispersive limit assumption cause the optimal gate implementation time to differ from the analytical expression above.  In the $n_\text{th} = 0$ case we see that the fidelity is not monotonic with Rabi detuning. This is explained by the competing effects of the dispersive assumption and the decoherence: Increasing the $| \Omega_R - \omega_m |/\lambda$ term first brings the system further to the dispersive regime. Beyond the optimal value of $| \Omega_R - \omega_m |/\lambda$, increasing the detuning makes the gate slower and the decoherence starts to dominate, therefore decreasing the gate fidelity. The optimal value of $| \Omega_R - \omega_m |/\lambda$ is monotonic in $\lambda$, which is in line with the interplay between the dispersive regime  condition and decoherence rate. In the $n_\text{th} = 1$ case, highest gate fidelities are achieved at higher Rabi detunings and $\lambda$. This is because the thermal population increases the Rabi detuning needed to obey the dispersive assumption.

\begin{figure}[t]
\captionsetup{justification=Justified}
\includegraphics[width=0.99\linewidth]{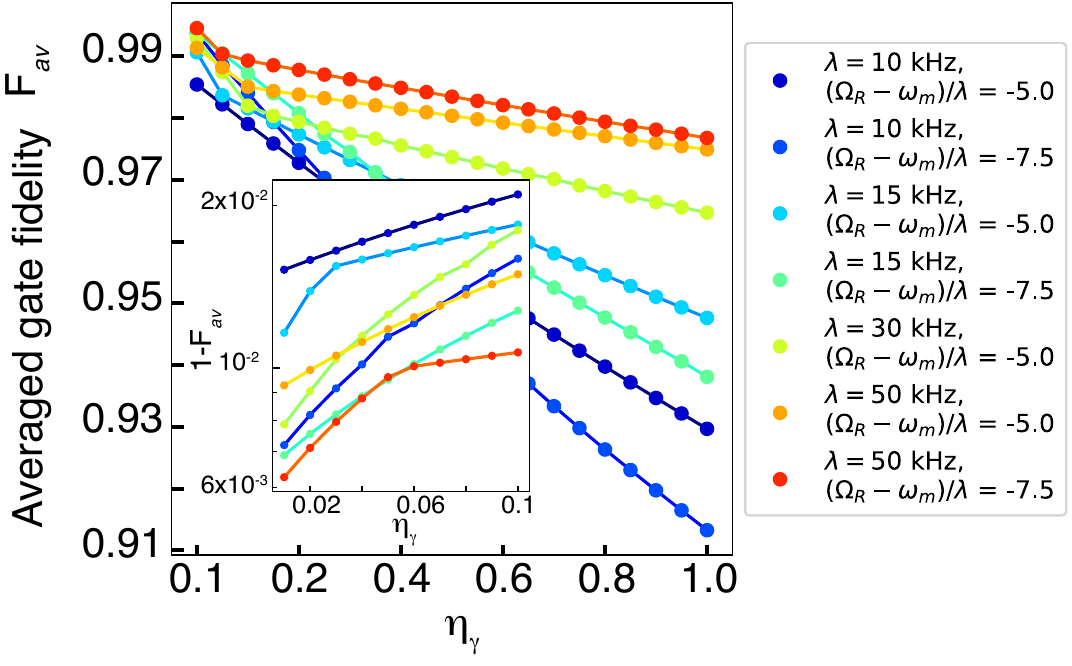}
\caption{The effect of decoherence on the $\sqrt{iSWAP}$-gate fidelity. Here, the horizontal axis is a front factor multiplying all the decay rates $\kappa_k$ in our master equation. $\eta_\gamma = 1$ corresponds to the qubit linewidths $1/T_1 = 10$ Hz, $1/T_2 = 100$ Hz, and mechanical linewidth of $100$ Hz used in our other simulations.}
\label{linewidth_gateFid_0exc}
\end{figure}

Figure \ref{gateFid_groundState} demonstrates the expected behaviour, but the fidelities reached for the $\sqrt{iSWAP}$ gate are not yet state-of-the-art. In order to reach higher gate fidelities we will need to improve on the decay rates of the systems. In Fig.~\ref{linewidth_gateFid_0exc}, we present the gate fidelity as function of the total amount of open system decoherence for selected parameter pairs $\lambda, (\Omega_R - \omega_m)$ from Fig.~\ref{gateFid_groundState}. Here, all the $\kappa_k$ decay rates are multiplied with the same coefficient $\eta_\gamma$. Here, $\eta_\gamma = 1$ corresponds to the linewidths used in the other simulations of this paper and $\eta_\gamma = 0$ would be the noiseless case. We see that for these parameter pairs, we can implement high-fidelity gate when the coupling strength $\lambda$ is sufficiently large and the qubit and mechanical resonator linewidths are small, and decreasing the linewidths has a large effect on the fidelity especially for smaller $\lambda$. 

An alternative way to entangle the qubits is by post-selection after measuring the shift of the mechanical resonance frequency caused by two qubits at the opposite ends of the nanobeam \cite{rosenfeld2021}. Also the microwave drive we apply to dress the qubits can be frequency modulated to implement the CNOT and iSWAP gates depending on how the modulation parameters are chosen \cite{srinivasa_cavity-mediated_2023}. These approaches are more complex to implement experimentally than the method we have numerically analyzed here, since the former depends on measuring the mechanical resonator and the latter relies on microwave modulation. Future work will show if the added complexity of these alternatives can lead to higher gate fidelities.

\section{Concrete implementation}\label{sec_siliconNanoBeams}

The specific implementation of the optomechanical quantum bus we propose is a modified version of the sliced silicon nanobeam introduced in Refs. \cite{leijssen_strong_2015,leijssen_nonlinear_2017,muhonen2019}. The design starts with a silicon nanobeam that works as a waveguide for light. A periodic hole pattern is introduced to the waveguide, and causes a spatially altering refracive index that basically forms a Bragg reflector (1D photonic crystal). A defect in the periodicity of the holes at the waveguide's center allows optical frequencies inside the Bragg mirror's bandgap travel there and thus the defect works as an optical cavity. 
To maximize the optomechanical coupling, the nanobeam is then sliced, leading to light concentration to the small gap \cite{leijssen_strong_2015}. The mechanical displacement from the nanobeam halves moving with respect to each other modifies the resonance frequency of the optical cavity. This causes modulation in optical phase quadrature at the mechanical frequency $\omega_m$, and therefore the interaction between the optical and mechanical modes is described by the standard optomechanical Hamiltonian in Eq.~\eqref{omLinH}. Isotopically purifying the top layer of the device will prolong the spin lifetime \cite{muhonen2014}, and in our experiments it has not been found to negatively impact the optical or mechanical properties of the devices (unpublished).

The mechanical resonance frequency of these devices is in the 1-10 MHz range, which allows bringing the dressed qubit into resonance with the mechanical resonator with reasonable microwave powers. The linewidth of the optical cavity is orders of magnitude larger than the mechanical frequency ($\sim$ 10 - 100 GHz). This means that the outgoing light information reflects the instantaneous mechanical position, which can be used for measurement based feedback cooling \cite{rossi_measurement-based_2018,guo_feedback_2019,kumar_single-laser_2022} and pulsed measurements \cite{vanner_pulsed_2011,vanner_cooling-by-measurement_2013,muhonen2019}. Feedback cooling will be needed to bring these low frequency resonators close to their quantum ground state whereas pulsed measurements allow detecting short-lived shifts in the mechanical frequency and give a single-shot spin readout method. 
\subsection{The spin-phonon coupling strengths}\label{Sec_couplMecs}

As outlined in section \ref{sec:dressedHam} the coherent spin-phonon coupling strength in the dressed frame is given by the change of the bare spin qubit resonance frequency by a displacement $x_{zpf}$. In our proposed platform, we consider two mechanisms to achieve this, both with their own strengths: motion-induced strain in the silicon lattice and a magnetic field gradient in the proximity of a micromagnet.

Both of these methods produce coupling strengths that vary heavily based on donor location, requiring precise control of donor placement for strong coupling. This can be achieved through deterministic ion implantation, where donors are placed in precisely determined locations using either a mask \cite{donkelaar_single_2015,jamieson_deterministic_2017} or an extremely precise focused ion beam \cite{pacheco_ion_2017}.

There are earlier demonstrations where strain has been used to couple solid-state defects to mechanical motion, e.g. in silicon carbide \cite{whiteley_spinphonon_2019}. Also an optomechanical system has been coupled to the electron spin of an NV center in diamond \cite{shandilya2021}. Micromagnets have been used to couple the electron spin in an NV center in diamond to the mechanical motion \cite{arcizet2011,pigeau2015}. Furthermore, resonantly driving an InAs quantum dot was shown to actuate the mechanical motion of a GaAs microwire via state-dependent strain \cite{kettler2021}. 
Periodically modulating strain at the spin location has been experimentally shown to drive the electron spin of an NV center in diamond using electrically controlled high-overtone bulk acoustic resonator (HBAR) \cite{macquarrie_mechanical_2013} and a piezoelectrically driven mechanical cantilever \cite{barfuss2015}.

Next, we will discuss these coupling mechanisms in our proposed quantum bus design and estimate with numerical simulations the spin-mechanics coupling strengths they can provide.

\subsubsection{Strain caused by the mechanical motion}

The mechanical displacement deforms the silicon's crystal lattice surrounding the donor atom, shifting the surrounding silicon atoms closer or further from it, thus changing the background charge environment of the loosely bound donor electron (the electron wavefunction covers several tens of silicon atoms). As a consequence, the electron's effective $g$ factor and the hyperfine interaction between the donor nucleus and electron change, resulting in a shift of the electron spin energy states. Until quite recently, the strain-induced shift was thought to be quadratic in strain as the valley repopulation theory \cite{wilson_electron_1961} predicts, but it was in 2018 experimentally shown to be linear for small strains \cite{mansir2018}. As the coupling strength is given by the strain that the donor experiences, the device geometry can be optimized by concetrating the amount of strain in a small region of the silicon crystal. 
The strain-induced change in the donor electron's spin resonance frequencies were reported to be 22 GHz/strain and 140 GHz/strain for phosphorus and bismuth donor in silicon, respectively\footnote{The reported values for the transitions/corresponding to the nuclear spin in its ground state and the 90 degree angle between the static magnetic field and the silicon crystal [001] direction.} \cite{mansir2018}. Strain is a very appealing choice for the coupling mechanism as it does not require any additional fabrication steps.

\begin{figure}
\centering
\includegraphics[width=0.95\linewidth]{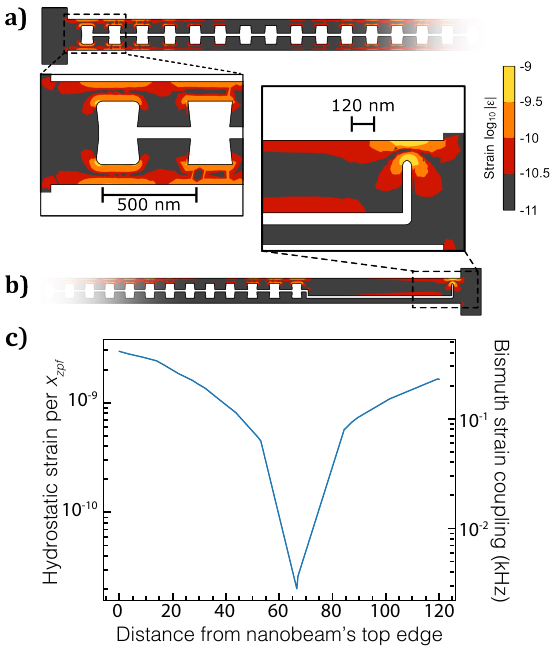}
\captionsetup{justification=Justified}
\caption{Strain induced coupling between the spin and the mechanical resonator. FEM simulation of the strain caused by displacement of $x_{xpf}$ for \textbf{a)} a conventional split nanobeam and \textbf{b)} a strain-optimized geometry. The insets show the high-strain regions at the ends of the nanobeams. The strain is more concentrated and five times higher in the optimized geometry where a vertical slit has been added. \textbf{c)} The coupling strength in the strain-optimized geometry: The left y-axis shows the strain along the vertical linecut at the pinch point and the right y-axis the corresponding spin-mechanics coupling strength for a bismuth donor electron spin.}
\label{strainComsol}
\end{figure}

Figure \ref{strainComsol} shows finite-element-method (FEM) simulations (using COMSOL software) of the strain distribution caused by a displacement with the magnitude of a zero point fluctuation amplitude ($x_{xpf}$) for two device geometries: in Fig.~\ref{strainComsol} \textbf{a)} similar to the devices used in \cite{leijssen_nonlinear_2017,leijssen_strong_2015,muhonen2019}, and in Fig.~\ref{strainComsol} \textbf{b)} the strain-optimized geometry. 
We see that in the conventional sliced nanobeam geometry the high-strain areas are widely spread near the ends and the center of the nanobeam. Importantly, the added vertical slit in the optimized geometry spatially confines and increases the amount of the strain, which leads to a 5-fold enhancement for the spin-mechanics coupling strength.
In the high-strain regions this optimized geometry provides coupling strengths of 60~Hz and 400~Hz for a single phosphorus and bismuth donor, respectively. The high-strain areas are near the narrow slits at both ends of the nanobeam, allowing use of the motion of the beam as a communication channel between two donor spins at the opposite ends as discussed in Sec.~\ref{sec_spinspin}.

\subsubsection{Magnetic field gradient of a micromagnet}

\begin{figure}
\captionsetup{justification=Justified}
\includegraphics[width=0.95\linewidth]{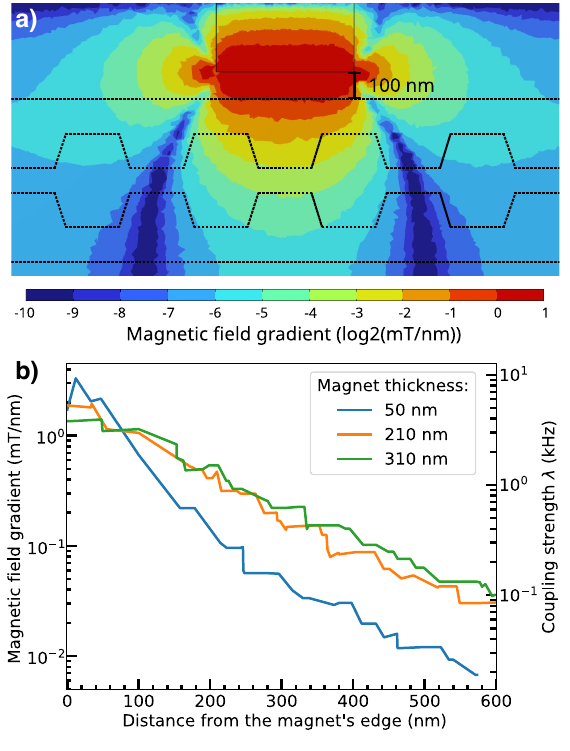}
\caption{Micromagnet induced magnetic field gradient for spin-mechanics coupling. \textbf{a)} COMSOL FEM simulation of the 2D spatial distribution of the magnetic field gradient caused by a micromagnet with dimensions of $500$ nm $\times~ 500$ nm $\times~ 300$ nm (length$\times$width$\times$height). The dotted line shows approximate position of the silicon nanobeam. \textbf{b)} COMSOL FEM simulation of the magnetic field gradient at the center of the nanobeam as function of the distance from the magnet's edge (left y-axis), and the resulting spin-mechanics coupling strength for a single phosphorus donor electron (right y-axis), where we have used $x_{zpf}$ = 100 fm and a square $500$ nm $\times~ 500$ nm magnet.
}
\label{magnetComsol}
\end{figure}

Alternatively, we can couple the spin-qubit to the nanobeam's mechanical motion via a magnetic field gradient induced by a stationary on-chip micromagnet fabricated near the nanobeam's center, where the beam's displacement is maximal. We present the COMSOL simulation of the magnetic field gradient for one example magnet geometry in Fig.~\ref{magnetComsol} \textbf{a)}. The simulation illustrates how the magnetic field gradient changes less drastically along the nanobeam, and thus the required donor implantation precision is set by the distance across the nanobeam.

When the nanobeam moves the distance $\Delta x$ towards the magnet, a donor in the beam experiences a magnetic field gradient $\nabla B$, modulating the Zeeman splitting of the electron by $\Delta x\nabla B \gamma_e$ which corresponds to the spin-mechanics coupling strength. The coupling rates for all donor species can thus be increased by maximizing both the distance traveled by the donor and the magnetic field gradient that the donor experiences. 

To maximize simultaneously both $\Delta x$ and $\nabla B$, the donor can be implanted at the center of the nanobeam where the beam's displacement is maximal, and the micromagnet should be placed near the donor, since $\nabla B$ is strongest near the magnet. Split silicon nanobeams with device geometries similar to ours have been reported to have $x_{zpf}$ values of 43~fm \cite{leijssen_nonlinear_2017} and 80~fm \cite{leijssen_strong_2015}. Still, $x_{zpf}$ can be enhanced with design changes: Lower mechanical resonance frequency should lead to floppier devices with higher $x_{zpf}$, as $x_{zpf}=\sqrt{\hbar / 2m_{eff}\omega_m}$, and lengthening the beam will decrease the mechanical resonance frequency $\omega_m$ more than it increases the effective mass $m_{eff}$. Furthermore, the magnetic field gradient $\nabla B$ can be increased by optimizing the magnet geometry, thickness, proximity, and material. The effects of proximity to $\nabla B$ and the resulting spin-mechanics coupling strength $\lambda$ for a single phosphorus donor are shown in Fig.~\ref{magnetComsol} \textbf{b)} for a selection of magnet thicknesses. We see that increasing the magnet thickness beyond 100 nm does not notably change the the magnetic field gradient.

\begin{figure}
\captionsetup{justification=Justified}
\includegraphics[width=0.8\linewidth]{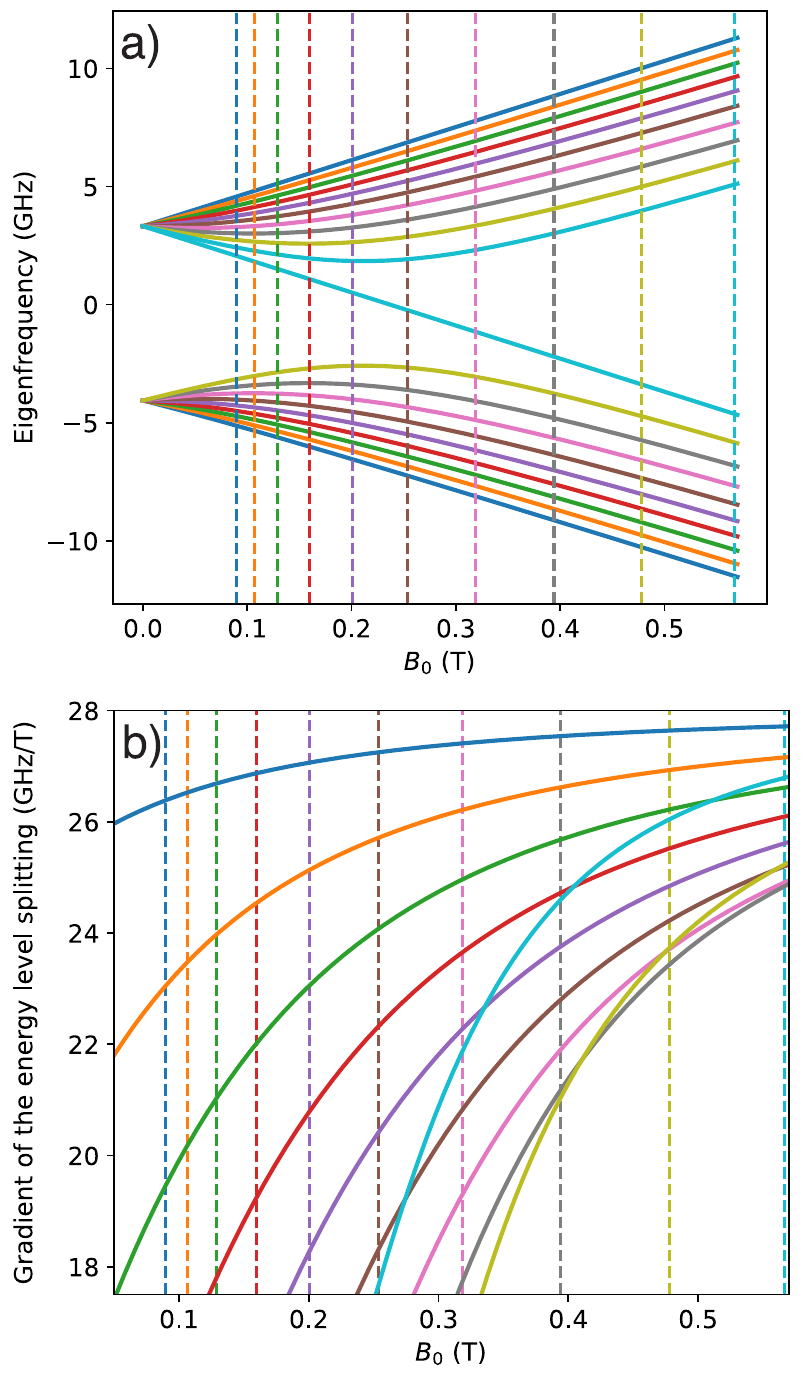}
\caption{\textbf{(a)} Bismuth donor electron energy level diagram. The dashed lines illustrate the magnetic fields corresponding to Zeeman splittings of 9.7 GHz between the matching colored energy levels. The lowest-field transition, provides the strongest spin-mechanics coupling via strain \cite{mansir2018}. \textbf{(b)} Gradient of the bismuth Zeeman splitting for all the nuclear spin states. The gradient varies a lot between the different transitions and it is not constant even at relatively high fields.}
\label{biTransitions}
\end{figure}

For a phosphorus donor in silicon, the Zeeman splitting obeys the linear relation of 28 GHz/T already at relatively small fields ($B\approx0.075~T$), making the relationship between $\nabla B$ and the coupling simple.
On the other hand, for bismuth, the large hyperfine constant and nuclear spin of 9/2 create a more complex energy level system, whose Zeeman splitting is nonlinear with respect to magnetic field also at relatively high magnetic fields, as shown in Fig.~\ref{biTransitions}. At the transition which is the most sensitive to changes of the external field $B$, the coupling strength for a bismuth donor is 7\% less than for a phosphorus donor and for the higher nuclear spin transitions it is even weaker. 

Therefore, the optimal choice of the donor species depends on the used coupling mechanism: while bismuth donors give rise to nearly ten-fold greater coupling strengths for strain coupling, phosphorus is superior for magnetic field gradient coupling at fields above $0.006~T$. We note that the coupling strength for a single phosphorus donor with magnetic field gradient coupling is an order of magnitude higher than for strain coupling for a single bismuth donor, making magnetic field gradient coupling with a phosphorus donor the better option for single spin readout. On the other hand, the area of active donor coupling is limited only to the center of the nanobeam and the inclusion of a metal micromagnet increases fabrication complexity.

\subsection{Practical considerations: Temperature dependence, ensemble coupling, and optical quality factor with magnet}

In the numerical modeling in Sections \ref{sec_readout} and \ref{sec_spinspin} we have concentrated on very low thermal phonon populations and single-spin coupling for practical numerical reasons. They demonstrate the interplay between the Rabi detuning and thermal phonon number that we must fulfill in order to rely on analytical estimates. One main message from them is that to reach the state-of-the-art two-qubit gate fidelity will require both ground state cooling and improvements in the linewidths whereas the spin readout using changes in the mechanical resonance frequency should be achievable with current structures, and possibly at elevated temperatures. 

A practical question for experiments is to ask at what temperature can the readout work with the above estimated coupling rates, for both single spins and ensemble of spins. To answer this, we can do some estimates based on the analytical formulae,
\(
\Delta \omega_m = \lambda^2 \Omega_R / [ 2(\omega_m^2 - \Omega_R^2) ],
\)
which can be written as
\begin{align}
    \Delta\omega_m &= \frac{\lambda^2\Omega_R}{2(\omega_m+\Omega_R)(\omega_m-\Omega_R)} %\nonumber \\
    %&= \frac{\lambda^2\Omega_R}{2(\Delta_R+2\Omega_R)(\Delta_R)} \nonumber \\
    %&= \frac{\lambda^2\Omega_R}{2(\Delta_R^2+2\Omega_R\Delta_R)} \nonumber \\
    = \frac{\lambda^2}{2\frac{\Delta_R^2}{\Omega_R}+4\Delta_R}
\end{align}

\begin{figure*}
    \centering
    \captionsetup{justification=Justified}
    \includegraphics[width=0.7\linewidth]{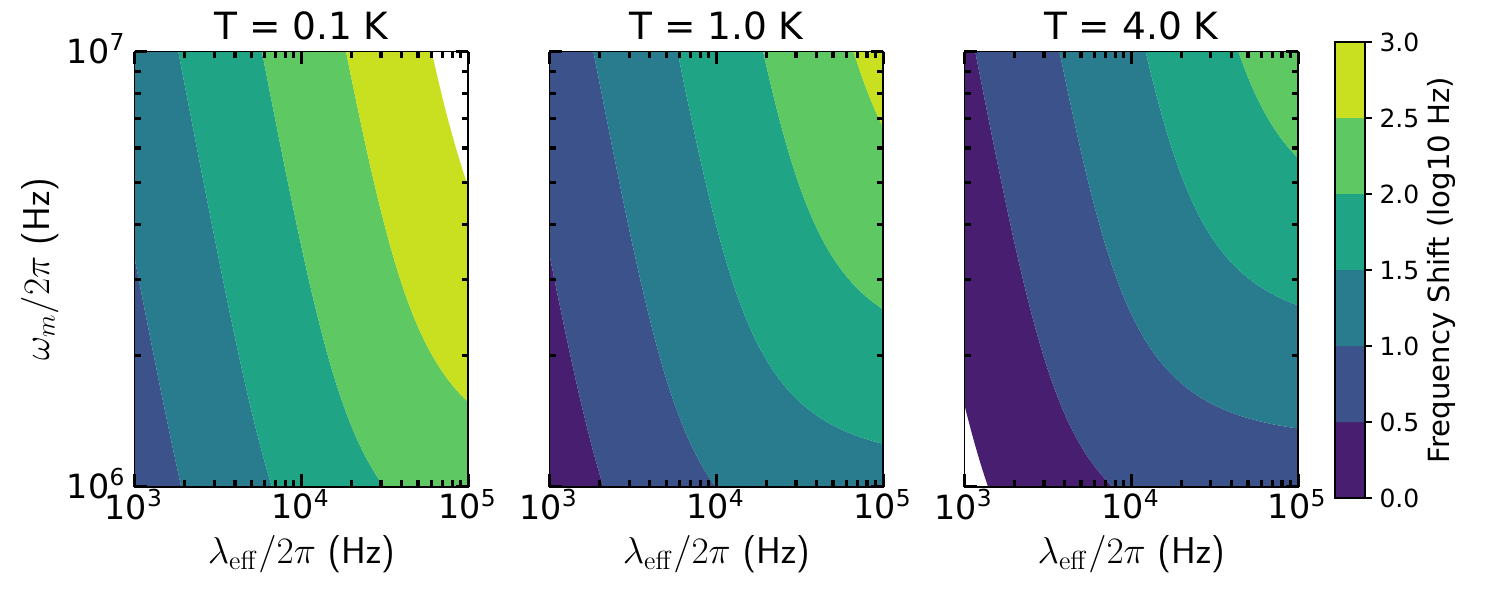}
    \caption{Frequency shift of the mechanical resonator as a function of the resonance frequency $\omega_m$ and coupling strength $\lambda$ at different bath temperatures.}
    \label{fig:readout_temperature}
\end{figure*}

Now, for numerical estimates we must assume something about the detuning. Below we will assume that $\Delta_R = \sqrt{n_{th}}\lambda $ and also that microwave driving is at resonance with the bare qubit frequency.
The calculated change in the resonance frequency as a function of $\omega_m$ and the coupling strength $\lambda$ at few different temperatures is shown in Fig. \ref{fig:readout_temperature}. To take a concrete example, let's consider a 3 MHz resonator, and assume we will need a 30 Hz shift, to be able to detect the shift. This would mean we need $\lambda = 3$ kHz at 0.1 K, $\lambda = 12$ kHz at 1.0 K and $\lambda = 50$ kHz at 4 K temperatures. Comparing to the COMSOL simulations above, the simulated coupling from the magnetic field gradient of a 310 nm thick magnet in Fig.~\ref{magnetComsol} shows that $\lambda$ = 3 kHz is reached with implantation precision of 50 nm from the nanobeam's top edge, which is within the reach of current ion implentation technologies. Thus, the magnetic field gradient coupling is strong enough for single-donor readout within the current fabrication capabilities. 
On the contrary, even for the optimized strain coupling device geometry the single-donor coupling strength is at most 0.4 kHz.

However, we can consider also the readout of an ensemble of donor spins confined in the same high-strain region or residing all within the magnetic field gradient. 
According to the Tavis-Cummings model, if the coupling strength between a qubit and a harmonic oscillator for a single qubit is $\lambda_1$, then the coupling strength between the ensemble spin of $n$ qubits and the harmonic oscillator is $\lambda_n = \sqrt{n}\lambda_\textrm
{RMS}$, where $\lambda_\textrm{RMS}$ is the root-mean-square of the individual coupling strengths of the donors. Note that the summation in square is crucial here for the strain coupling as otherwise donors in opposite sides of the beam would cancel out in the calculation.

If we now assume an implantation area of $100\times100\times200$ nm$^3$ (realistic based on the simulations above and current implantation accuracies with focused ion beams) and implantation density of 1e17 cm$^{-1}$, we will have 220 donors in this area. This would mean we need $\lambda_\textrm{RMS} = 0.22$ kHz at 0.1 K, $\lambda_\textrm{RMS} = 0.8$ kHz at 1.0 K and $\lambda_\textrm{RMS} = 3.1$ kHz at 4 K temperatures, which means that ensemble readout is possible at 4 K with the magnetic field gradient coupling and at 0.1 K with strain coupling. Notably, we only considered one implantation site here, even though there are in principle multiple strain areas in a single beam, increasing the signal.

\begin{figure}
    \captionsetup{justification=Justified}
    \includegraphics[width=0.8\linewidth]{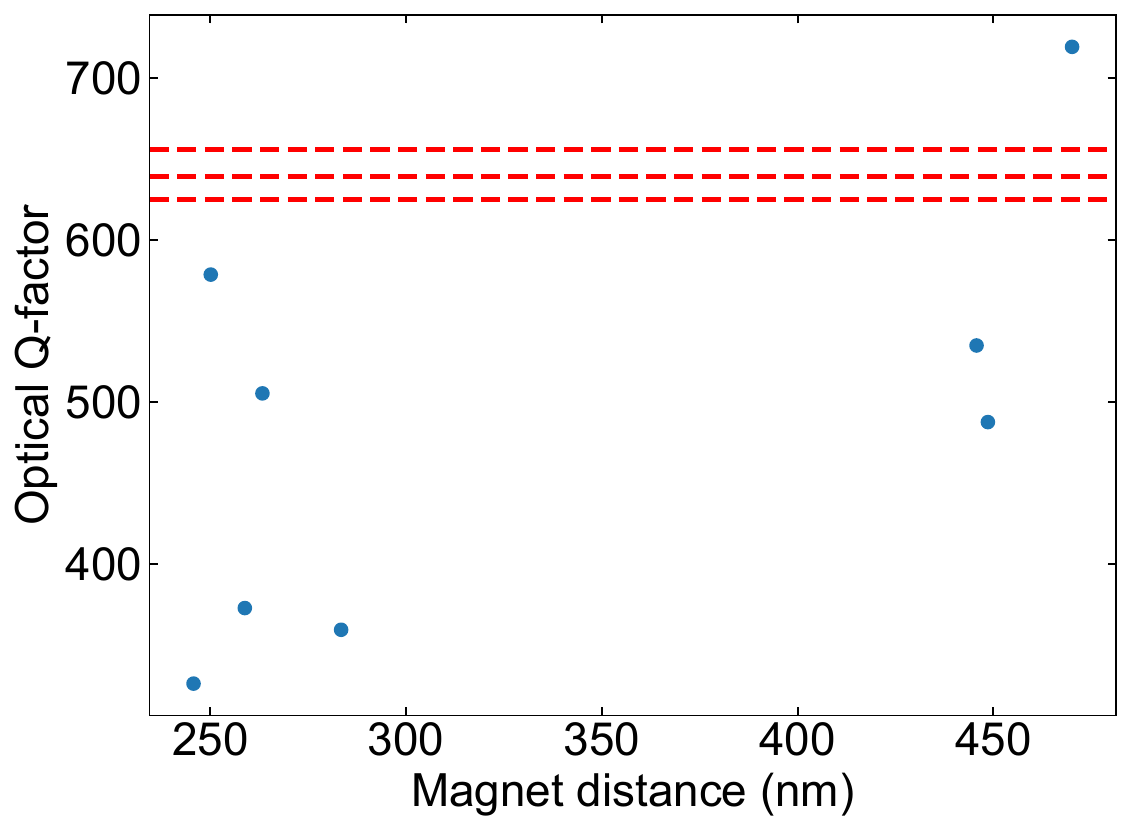}
    \caption{The optical Q-factors ($\omega_c/\kappa$) of optomechanical resonators as a function of the closest distance between beam and magnet (blue dots) as well as of magnet-free control samples (red dashed lines). Here the control samples optical quality was not as high as in previous demonstration due to non-ideal lithography. This still shows a definite lower bound for the achievable optical quality factor with the micromagnet.}
    \label{opticalQ}
\end{figure}

Concerning the optomechanical readout of the spin-qubits, it should be noted that metal magnets have a large imaginary component to their refractive indices, leading to significant absorption. If this impacts the optical cavity mode notably, it could be detrimental to the optical readout. To test this, we have experimentally fabricated a series of optomechanical cavities in close proximity to nickel micromagnets and performed measurements to characterize their optical and mechanical resonances. There were no differences in the mechanical resonance between the magnet and control samples, nor in the optical resonance frequency $\omega_c$. The presence of the micromagnet increased the optical linewidth $\kappa$ and thus decreased the optical quality factor. The resonators that were closest to the magnet experienced the sharpest drop in quality, as seen in Fig.~\ref{opticalQ}. Importantly, the optical cavity still survived the presence of the magnet and thus the micromagnet does not critically compromise the performance of the quantum bus. The observed Q-values were enough for high-quality optomechanical interface.

%\textbf{Discussion of the performance for 2-qubit gates pending.}\\

\section{Conclusions}

In summary, we have introduced and quantitatively analyzed an optomechanical quantum bus that enables both long-range coupling and optical readout of donor spin qubits in silicon. By exploiting mechanically mediated interactions in a microwave-dressed frame, the proposed architecture circumvents the stringent placement requirements and wiring overhead that have so far limited the scalability of donor-based quantum processors. The same mechanical degree of freedom provides a dispersive, quantum non-demolition readout channel that is compatible with telecom-wavelength photonics and does not require the mechanical resonator to be in its ground state.

We have numerically analyzed the experimentally relevant regimes for both readout and entangling operations. While achieving state-of-the-art two-qubit gate fidelities will require further reductions in dressed spin and mechanical decoherence, the requirements for spin readout are already compatible with existing optomechanical devices and deterministic donor implantation techniques. In particular, magnetic field gradient coupling enables single-donor readout within current fabrication tolerances, whereas ensemble coupling offers a viable route toward elevated-temperature operation.

More generally, this work establishes mechanical resonators as a versatile quantum resource for donor spin qubits, enabling controlled spin–spin interactions over mesoscopic distances and direct interfacing with optical fields. The proposed platform thus provides a concrete pathway toward modular, silicon-based quantum architectures and hybrid spin–phonon–photon systems, bridging solid-state qubits with photonic quantum networks in a fully CMOS-compatible material system. Ultimately, also coherent coupling from the spins to photons could be considered.

\section*{Acknowledgements}
This project has received funding from the European Research Council (ERC) under the European Union’s Horizon 2020 research and innovation program (Grant Agreement No. 852428), from Research Council of Finland (Grant No. 321416 and 354735), and from the Finnish Quantum Flagship (project no. 359240, University of Jyväskylä).

\bibliography{references}

\end{document}